\documentclass{article} % For LaTeX2e
\usepackage{iclr2021_conference,times}

\usepackage{amsmath,amsfonts,bm}

\def\eqref#1{equation~\ref{#1}}
\def\1{\bm{1}}

\DeclareMathAlphabet{\mathsfit}{\encodingdefault}{\sfdefault}{m}{sl}
\SetMathAlphabet{\mathsfit}{bold}{\encodingdefault}{\sfdefault}{bx}{n}

\usepackage{graphicx}
\usepackage{hyperref}
\usepackage{booktabs}
\usepackage{multirow}
\usepackage{url}
\usepackage{wrapfig}
\usepackage{amsmath}
\usepackage{lipsum}
\usepackage{makecell}
\usepackage{enumitem}
\title{robust audio anomaly detection}

\author{Wo Jae Lee\thanks{Equally contributing authors.} \thanks{Work conducted while the author was at Amazon Web Services.}  \\
Purdue University\\
\And
Karim Helwani\footnotemark[1] , Srikanth Tenneti, \& Arvindh Krishnaswamy\\
Amazon Web Services \\
\texttt{\{helwk\}@amazon.com}
}

\iclrfinalcopy 
\begin{document}

\maketitle

\begin{abstract}
We propose an outlier robust multivariate time series model which can be used for detecting previously unseen anomalous sounds based on noisy training data. The presented approach doesn't assume the presence of labeled anomalies in the training dataset and uses a novel deep neural network architecture to learn the temporal dynamics of the multivariate time series at multiple resolutions while being robust to contaminations in the training dataset. The temporal dynamics are modeled using recurrent layers augmented with attention mechanism. These recurrent layers are built on top of convolutional layers allowing the network to extract features at multiple resolutions. The output of the network is an outlier robust probability density function modeling the conditional probability of future samples given the time series history. %We further discuss synthetic augmentation approaches tailored for multivariate time series forecasting and anomaly detection.
State-of-the-art approaches using other multiresolution architectures are contrasted with our proposed approach. We validate our solution using publicly available machine sound datasets. We demonstrate the effectiveness of our approach in anomaly detection by comparing against several state-of-the-art models.
\end{abstract}

\section{Introduction}
The task of anomaly detection has many natural applications and has been studied within diverse research areas and application domains such as monitoring financial indicators, machine health, cloud resources, etc.  \citep{ayed2020anomaly}. For example, anomaly detection methods are applied to maximize system uptime and to improve operational performance in manufacturing plants by detecting abnormal behaviors (e.g., breakdown) of manufacturing equipment. A concise overview of anomaly detection approaches has been presented in \citet{chandola2009anomaly}. A large class of anomaly detection methods assume the existence of only normal data for training. Hence, the main challenge in designing a reliable anomaly detector is to find the best representation of the data which neither generalizes to unseen anomalies resulting in false negatives, nor overfits on existing data and hence, recognize falsely unseen normal data points as anomalies which results in alerting the user too often.

%\subsection{Contribution}
In a realistic setting, it is often challenging to obtain a clean training dataset and it is common to encounter outliers in the training dataset. For example, time series data collected by vibration sensors or microphones in an industrial environment is typically noisy as a result of dynamic interaction of multiple factors. Therefore, a mixture of, generally, non-stationary signals from multiple unknown sources is added to the signal of interest. The limitation of current anomaly detection methods is their lack of ability to deal with contaminated training datasets while capturing temporal dependencies across multiple time steps. In \citet{nalisnick2018deep}, the limitations of generative models on the anomaly detection / density estimation are discussed, and the necessity of the study on the robust to  out-of-distribution inputs are emphasized. The present approach targets this limitation explicitly as it develops a robust multivariate time series anomaly detection model, in which we evaluate the probability of current samples given the time series history using a  conditional probability density function learned from the training data. Three key characteristics of our method are: 1) The ability to model temporal dependency in the multivariate time-series data, 2) The ability to track the signal dynamics at multiple resolutions, and 3) Robustness to outliers and noisy training datasets. Our approach is based on an attention mechanism augmented recurrent neural network (RNN) architecture which captures the temporal dependencies in the multivariate time series. Further, we propose a hierarchical structure to allow the extraction of global features (multiscale resolution), and finally, our approach models the time series innovation with an outliers robust probability distributions. %

%\subsection{Related Prior Work}
In prior works, anomaly detection methods have been discussed in settings where it is assumed that the input data is either independent and identically distributed or in explicit time series settings, where the temporal dynamic of the time series is explicitly modeled, e.g., using an autoregressive model as it is often the case in traditional change point detection methods \citep{gustafsson2000adaptive}. In both mentioned cases, approaches to solve the anomaly detection problem are broadly either \textit{Generative} or \textit{Discriminative} \citep{wu2019tale}.

In the generative approaches, a model is trained to generate samples of the data based on a certain latent representation. If the actual incoming measurement samples and the generated samples differ beyond a tolerable threshold, an anomalous data point is identified. Generative models include the popular autoencoder approaches \citep{zhou2017anomaly,zong2018deep,gong2019memorizing} and the methods based on the recent neural density estimation \citep{papamakarios2017masked,germain2015made,rezende2015variational} such as \citet{yamaguchi2019adaflow,6679866,Giri2020gmade}. Shortcomings of the density estimation based anomaly detectors have been discussed empirically in \citet{nalisnick2018deep}. They argued that a well-calibrated generative model is not able to assign higher probability to the training data than to some other out-of-distribution data. These shortcomings make the potential power of discriminative approaches apparent.   

In the lack of anomalous data, the development of discriminative approaches employs the synthetic generation of anomalous data as perturbations on the normal data, e.g., \citet{Giri2020}. Clearly, the generation of synthetic anomalous data is problem specific and being able to generate relevant anomalous data points is often not a given. %the data augmented through linear %transformation can play as a %regularization effect \cite{wu2020}  
In this study, we argue that with generative models state-of-the-art anomaly detection performance can be achieved, if the target distribution is chosen to be outlier robust. Intuitively, a well performing anomaly detector should excel at modelling the boundaries of the normal data distribution and hence, it models the probability distribution of the data "tightly" which can be achieved by outlier robust modeling.

A popular representative of time series methods is the autoregressive model. In this model, the next sample is predicted as a linear function of previous samples while modeling the uncertainty of the model output as allowable \textit{innovation} of the system \citep{kailath1968innovations}. For example, in a multivariate linear prediction setting, often a convolutive model is used where the innovation is modeled as
\begin{align}
    \mathbf{e}(t) &=\mathbf{x}(t+1)-{\mathbf{W}}^{T}(t) \mathbf{X}(t),
\end{align}
where $\mathbf{X}(t) =\left[\mathbf{x}_{1}^{T}(t), \mathbf{x}_{2}^{T}(t), \cdots, \mathbf{x}_{P}^{T}(t)\right]^{T}$, $\mathbf{x}_{p}(t) =\left[x_{p}(t), x_{p}(t-1), \cdots, x_{p}(t-L+1)\right]^{T}$, $\mathbf{x}(t) =\left[x_{1}(t), x_{2}(t), \cdots, x_{P}(t)\right]^{T}$,  $\{ \cdot \}^T$ denotes transpose, $L$ is the prediction context, $p$ is the feature index, and $t$ is the time index. The estimation of the model parameters can be done using least squares approach which implies a Gaussian innovation model.

Time series anomaly detection using recurrent architecture has been proposed in \citet{marchi2015non} with application to audio novelty detection where the auditory spectral features are processed by an autoencoder. %This approach relies on the reconstruction error which the denoising autoencoder commits trying to predict and reconstruct a novel sound which the network has never seen in the training phase. 
More recently, a U-Net based architecture with convolutional LSTM layers has been proposed in \citet{zhang2019deep}. This approach is also temporally multiresolutional as it transforms the input into correlation matrices of multiple time lags. %The network is trained to reconstruct the input features from a compressed representation with minimum least squares error without explicit modeling of the probabilty of current samples given the time series history. 
Methods for recurrent density estimation have been proposed in \citet{oliva2017recurrent}, where the power of gated-RNNs has been highlighted for this task as such models were able to scan through previously seen dimensions remembering and forgetting information as needed for conditional densities without making any strong Markovian assumptions. Similar architecture for forecasting has been presented in \citet{salinas2020deepar}.%, . 
\section{Predictive Robust Parametric Density Model}\label{model_archi}
\begin{wrapfigure}{r}{0.4\textwidth}
\begin{center}
	\includegraphics[width=0.37\textwidth]{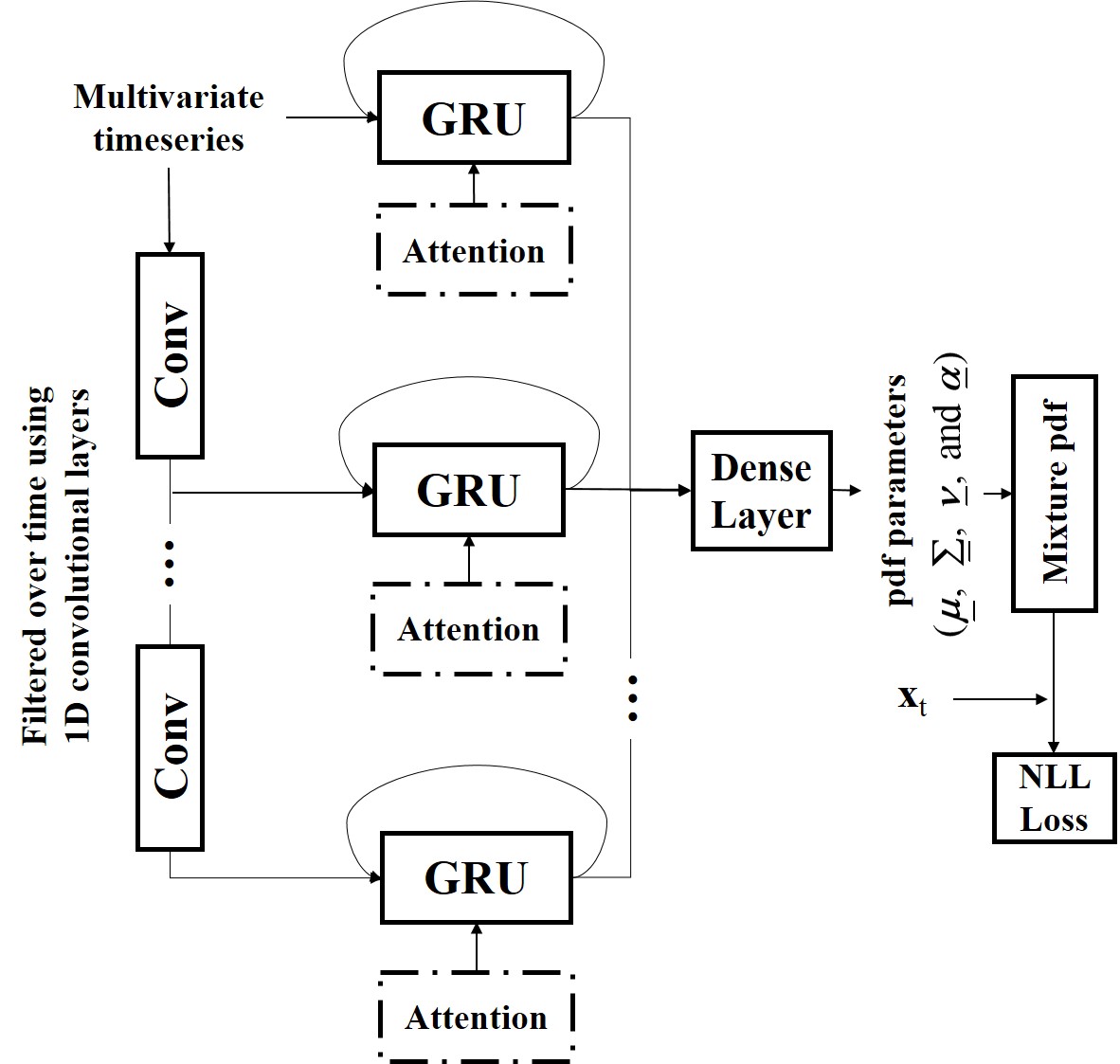}
	%\vspace{-10pt}
	\caption{Architecture of the proposed anomaly detection system.}
	\label{fig1}
\end{center}
\end{wrapfigure}

An overview of the model architecture we are proposing is shown in Fig~\ref{fig1}. The model aims at predicting a probability density function (pdf) of a next point given its past using a recurrent architecture where a conditional distribution is modeled as a function of past points using hidden state information $\mathbf{h}_{t}$ as follows
\begin{align}\nonumber
P\left(\mathbf{x}_{t} \mid \mathbf{x}_{t-1, \ldots, 0}\right)\approx P\left(\mathbf{x}_{t} \mid f\left(\mathbf{x}_{t-1, \ldots,0}\right)\right)=:P\left(\mathbf{x}_{t} \mid \mathbf{h}_{t}\right).
\end{align}
To be robust to outliers during training, we include a parametric estimation of the conditional probability using a robust probability density function. %
\citet{hampel1971general} gave a general formulation of distributional robustness in terms of a stability principle. Thereby, the deviations of a certain test statistic should be bounded for bounded deviations of the stochastic model. Consequently, a general approach to obtain an optimum robust estimator is based on the minimax principle \citep{huber1981robust}.
Generally, the minimax approach aims at the least favorable situation for which it suggests the best solution. Hence, in our context, the basic underlying idea for a robust pdf for the next sample is to find a worst case pdf within the given class of models with a certain $\epsilon$-deviation. As a suitable definition of the $\epsilon$-deviation, a typically  introduced model is the $\epsilon$-contamination model 
\begin{align}\label{eq:contamination}
p(\mathbf{y})=(1-\epsilon) p_0(\mathbf{y})+\epsilon p_{\text {outliers }}(\mathbf{y}),
\end{align}
where $p_0(\mathbf{y})$ denotes the distribution of the clean data, $p_{\text {outliers }}(\mathbf{y})$ the distribution of the outliers in the training data, and $\epsilon$ is the contamination parameter. %. 
As shown in \cite{huber1981robust}, considering the maximum asymptotic variance of the estimates for all possible densities within the assumed $\epsilon$-deviation yields a powerful approach to optimum outlier robustness.\\
Student $t$ distribution can be less sensitive to outliers than Gaussian distribution by controlling the degree of freedom parameter \citep{peel2000robust,svensen2005robust,wong2009student}. \\
In this paper, we will show how the Student $t$ mixture model can be seen as a special case of the robust distribution framework presented. The parametric model we assume is
\begin{align}\label{eq:9}
P(\mathbf{x}_{t}|\mathbf{x}_{t-1:0} ; \boldsymbol{\theta})=\sum_{i=1}^{c} \alpha_{i} f\left(\mathbf{x}_{t}|\mathbf{x}_{t-1:0} ; \boldsymbol{\mu}_i, \boldsymbol{\Sigma}_{i}, \nu_{i}\right),
\end{align}
where $\alpha_i$ denoting the responsibility, $c$ the number of components, and  
\begin{align}
f(\mathbf{y}; \boldsymbol{\theta}=(\boldsymbol{\mu}, \boldsymbol{\Sigma}, \nu))=\frac{\Gamma((\nu+P) / 2)}{\Gamma(\nu / 2) \nu^{P / 2} \pi^{P / 2}|\boldsymbol{\Sigma}|^{1 / 2}}\nonumber%\\
\cdot\left[1+\frac{1}{\nu}(\mathbf{y}-\boldsymbol{\mu})^{\mathrm{T}} \boldsymbol{\Sigma}^{-1}(\mathbf{y}-\boldsymbol{\mu})\right]^{-(\nu+P) / 2},\nonumber
\end{align}
where $\boldsymbol{\mu}$ representing the mean, $\boldsymbol{\Sigma}$ the covariance matrix, $\nu$ the degree of freedom, $P$ the dimension, and $\Gamma(\cdot)$ is the Gamma function. It is known that the Student $t$ distribution can be expressed as a scale mixture model \citep{murphy2012machine}
\begin{align}
\mathcal{T}\left(\mathbf{x}_{i} \mid \boldsymbol{\mu}, \mathbf{\Sigma}, \nu\right)=\int \mathcal{N}\left(\mathbf{x}_{i} \mid \boldsymbol{\mu}, \mathbf{\Sigma} / z_{i}\right) \operatorname{Ga}\left(z_{i} \mid \frac{\nu}{2}, \frac{\nu}{2}\right) d z_{i}, \nonumber
\end{align}
with $\operatorname{Ga}\left(z_{i} \mid \frac{\nu}{2}, \frac{\nu}{2}\right)$ being the gamma distribution with the shape and rate parameters being equal $\frac{\nu}{2}$. 
Splitting up the integral around the $z=1$ offers a representation similar to Eq.~(\ref{eq:contamination}). For the estimation of the parameters of the assumed model, we use a network inspired by the so-called WaveRNN \citep{kalchbrenner2018efficient} and augment this architecture with an attention mechanism as a temporal regularizer as we describe in Appendix \ref{TAM}, and a multiresolutional architecture employing convolutional layers described in Appendix \ref{MA}. These two architectural features help the network finding relevant patterns in long sequential data and allow the extraction of global features. The basic core operations of the network at a given resolution are as follows% %
\begin{gather*}
\mathbf{\check{x}}_{t} =\left[\mathbf{x}_{t-1:0} ; \mathbf{c}_t\right], \quad \mathbf{u}_{t} =\sigma\left(\mathbf{W}_{u} \mathbf{h}_{t-1}+\mathbf{U}_{u} \mathbf{\check{x}}_{t}\right), \quad \mathbf{r}_{t} =\sigma\left(\mathbf{W}_{r} \mathbf{h}_{t-1}+\mathbf{U}_{r} \mathbf{\check{x}}_{t}\right),\\
\tilde{\mathbf{h}}_{t} =\tanh \left(\mathbf{r}_{t} \circ\left(\mathbf{W}_{h} \mathbf{h}_{t-1}\right)+\mathbf{U}_{h} \mathbf{\check{x}}_{t}\right), \quad
\mathbf{h}_{t} =\mathbf{u}_{t} \circ \mathbf{h}_{t-1}+\left(1-\mathbf{u}_{t}\right) \circ \tilde{\mathbf{h}}_{t}, \\
\underline{\boldsymbol{\mu}}_{t}=\mathbf{W}_{\mu} \operatorname{relu}\left(\mathbf{W}_{1} \mathbf{h}_{t}\right), \quad \underline{{\nu}}_{t}=\operatorname{\sigma^{+}}\left(\mathbf{W}_{\nu} \operatorname{relu}\left(\mathbf{W}_{1} \mathbf{h}_{t}\right)\right), \quad \underline{{\alpha}}_{t}=\operatorname{softmax}\left(\mathbf{W}_{\alpha} \operatorname{relu}\left(\mathbf{W}_{1} \mathbf{h}_{t}\right)\right),\\\nonumber
\underline{\boldsymbol{\Sigma}}_{\textrm{diag}, t} =\operatorname{softplus}\left(\mathbf{W}_{2, \Sigma} \operatorname{relu}\left(\mathbf{W}_{1} \mathbf{h}_{t}\right)\right), \quad \underline{\boldsymbol{\Sigma}}_{\textrm{lower}, t} =\mathbf{W}_{1, \Sigma} \operatorname{relu}\left(\mathbf{W}_{1} \mathbf{h}_{t}\right),
\end{gather*}%\label{eq:11}%

where the $\mathbf{W}_{(\cdot)}$
and $\mathbf{U}_{(\cdot)}$ matrices are the GRU weights, $\mathbf{c}_t$ are conditioning parameters, e.g., features extracted as multiple resolutions, $\mathbf{u}_{t}$ is the update gate vector, $\mathbf{r}_{t}$ is the reset gate vector,  $\circ$ denotes the Hadamard product, the underlined quantities denote a concatenation of the mixture components parameters, $\boldsymbol{\Sigma}_{\textrm{lower}} $ denotes the components in the lower triangular part of the covariance matrix, $\boldsymbol{\Sigma}_{\textrm{diag}}$ stands for the diagonal components of the covariance matrix, $\operatorname{\sigma}$ is the sigmoid function, $\operatorname{\sigma^{+}}$ is a scale sigmoid function,  and the matrices $\mathbf{W}_{\nu,\{1,2\}\Sigma,\mu,\alpha}$ are matrices with block zero matrices selecting the respective output channels of the last layer. Here, we omit the biases for clarity. %. 
During the training phase, negative log likelihood (NLL) is adopted as a loss function, which can be written as follows
\begin{align}\label{eq:12}
L(\mathbf{x}_{t}|\mathbf{x}_{t-1:0})=-\log (P(\mathbf{x}_{t}|\mathbf{x}_{t-1:0};  \boldsymbol{\theta}))
=-\log \left(\sum_{i=1}^{c} \alpha_{i} f\left(\mathbf{x}_{t}|\mathbf{x}_{t-1:0}; \boldsymbol{\mu}_{i}, \boldsymbol{\Sigma}_{i}, {\nu}_{i}\right)\right),
\end{align}
where $f(\cdot)$ is the assumed innovation distribution as discussed above.
\begin{table}[t]
\caption{Overall average performance of the proposed system (RSMM-MR) measured in AUC / pAUC and comparison to the baseline models.}
\begin{center}
\begin{small}
\begin{sc}
\
\resizebox{1\textwidth}{!}{%
\begin{tabular}{|c|c||c|c|c|c|c|c|}
\toprule
\multicolumn{2}{|c||}{\multirow{2}{*}{\textbf{Method}}}          & \multicolumn{4}{c|}{\textbf{MIMII}}                                            & \multicolumn{2}{c|}{\textbf{ToyADMOS}}     \\ \cline{3-8} 
\multicolumn{2}{|c||}{}                                 & \textbf{Fan}             & \textbf{Pump}            & \textbf{Slider}          & \textbf{Valve}           & \textbf{Toy Conveyor}    & \textbf{Toy Car}         \\ \hline

\multirow{12}{*}{\begin{tabular}[c]{@{}c@{}}Solo\\ Models\end{tabular}}    & OC-SVM                     & 0.6455 / 0.5824 &0.6439 / 0.5679&	0.5769 / 0.5029&	0.4609 / 0.4970&0.5295 / 0.5087&0.6189 / 0.5318\\\cline{2-8} 
                          & IF                         &0.6816 / 0.5374 &0.6960 / 0.5623&	0.5777 / 0.5029&	0.4640 / 0.4952&0.5819 / 0.5125&0.6941 / 0.5429 \\ \cline{2-8} 
                          & AE                         & 0.7878 / 0.6466 &0.7947 / 0.6280&	0.6375 / 0.5190&	0.5242 / 0.5067&0.8066 / 0.6432&0.7955 / 0.6204 \\ \cline{2-8} 
                          & GMM                        &0.8257 / 0.6719 &0.8684 / 0.6707&	0.7533 / 0.5635&	0.5678 / 0.5194& 0.8243 / 0.6636& 0.8017 / 0.6254 \\ \cline{2-8} 
                          & MSCRED                      &0.5575 / 0.5367 &0.4940 / 0.5148&	0.4696 / 0.4901&	0.4972 / 0.4984 & 0.4790 / 0.4845 & 0.4304 / 0.4984 \\ \cline{2-8} 
                          & DAGMM                        &0.6438 / 0.5364 &0.6996 / 0.5907 &	0.6492 / 0.5473 &	0.4729 / 0.5021& 0.6223/  0.5181& 0.6730 / 0.5644 \\ \cline{2-8} 
                          & DCASE Baseline         & 0.8280 / 0.6579	&	0.8237 / 0.6411&0.7941 / 0.5906&0.5737 / 0.5079&0.8536 / 0.6495	&0.8014 / 0.6617 \\ \cline{2-8} 
                          & RGMM                       & 0.8206 / 0.6504	&	0.8455 / 0.6590&0.9174 / 0.6909&0.8815 / 0.7263&0.8195 / 0.6497	&0.8663 / 0.7277 \\ \cline{2-8} 
                          & RGMM-MR                    & 0.8264 / 0.6479&0.8702 / 0.6646&0.9370 / 0.7429&0.9572 / 0.8615	&0.8660 / 0.6769	&0.8812 / 0.7410 \\ \cline{2-8} 
                          & RSMM                       & 0.8560 / 0.6804&0.8571 / 0.6564&0.9358 / 0.7788&0.8672 / 0.7041&	0.8713 / 0.6832&0.8805 / 0.7501 \\ \cline{2-8} 
                          & RSMM-MR-W/o Attention           & 0.8783 / 0.7013 & 0.8921 / 0.7082&  0.9711 / 0.8546 & 0.9531 / 0.8056 &	0.9036 / 0.7168 & 0.8770 / 0.7518 \\ \cline{2-8} 
                          & \textit{RSMM-MR}                  & \textbf{\textit{0.8797}} / \textbf{\textit{0.7054}} & \textbf{\textit{0.9096}} / \textbf{\textit{0.7353}}& \textbf{\textit{0.9819}} / \textbf{\textit{0.9052}}& \textbf{\textit{0.9752}} / \textbf{\textit{0.8873}} &	\textbf{\textit{0.9316}} / \textbf{\textit{0.7875}} & \textbf{\textit{0.9198}} / \textbf{\textit{0.8112}} \\
\bottomrule\bottomrule
\multirow{2}{*}{\begin{tabular}[c]{@{}c@{}}Ensemble \\ Models\end{tabular}} & DCASE winner             & 0.9454 / 0.8430 & 0.9365 / 0.8173&	0.9763 / 0.8973&	0.9613 / 0.9089&0.9119 / 0.7334& 0.9434 / 0.8973 \\ \cline{2-8} 
                          & RSMM-MR with DCASE winner & \textbf{0.9891} / \textbf{0.9503}& \textbf{0.9590} / \textbf{0.8817}& \textbf{0.9979} / \textbf{0.9928} & \textbf{0.9849} / \textbf{0.9554} &	\textbf{0.9316} / \textbf{0.7876} & \textbf{0.9583} / \textbf{0.9262} \\ 
\bottomrule

\end{tabular}
 	}
	\label{table1_all}
	\end{sc}
\end{small}
\end{center}
\end{table}

\section{Experimental Results}
The datasets used for evaluating the presented approach are the same used for DCASE challenge task 2 \citep{urldcase}. The desciptions of the dataset, data preprocessing, and evaluation metrics can be found in Appendix \ref{dataset}.
We compare the performance of the proposed model with eight baseline models including the DCASE winner \citep{Giri2020}, the DCASE baseline \citep{Koizumi_DCASE2020_01}, two state-of-the-arts anomaly detection models (MSCRED \citep{zhang2019deep} and DAGMM \citep{zong2018deep}), and four traditional anomaly detection models. The traditional models we chose are One-class Support Vector Machine (OC-SVM), Isolation Forest (IF), Autoencoder (AE), and Gaussian Mixture Model (GMM). Each of the baseline models is described in detail in Appendix \ref{base}. We also consider the four variants of the proposed model for ablation study with respect to multi-resolution, attention mechanism, and Student $t$ mixture model. The four variants are named as follow: (1) RGMM: we don't implement convolutional layers and hence, exclude the multi-resolutional architecture and replace a mixture of multivariate Student $t$ with a mixture of multivariate Gaussian distribution, (2) RGMM-MR: we implement a multiresolutional architecture but replace a mixture of multivariate Student $t$ with a mixture of multivariate Gaussian distribution, (3) RSMM: we don't implement the convolutional layers but use a mixture of multivariate Student $t$ to parameterize the target pdf, and (4) RSMM-MR-W/o Attention: we don't implement the temporal attention mechanism in the proposed model.

The details of the model implementation is presented in Appendix E. The results of the experiment are summarized in Table~\ref{table1_all}. Additional descriptions including the results on each machine IDs are also reported in Appendix \ref{allexp}. RSMM-MR outperforms most of the baseline models, and shows the comparable results to the DCASE winner although the DCASE winner is an ensemble of multiple anomaly detectors. We further show the results that can be obtained by ensembling the presented RSMM-MR with the system described in \cite{Giri2020} (RSMM-MR with DCASE Winner in Table \ref{table1_all}) and show the achieved superior performance of the obtained ensemble. Further, we conducted ablation test with respect to multi-resolution, attention mechanism, and Student $t$ mixture model. The results of the four variants demonstrate the benefit of the individual components in the proposed system. Quantitative ablation analysis is given in Appendix \ref{abal}. Further, an analysis on the robustness to synthetically added Gaussian noise bursts to training data is provided in Appendix \ref{robustness}.

\bibliography{iclr2021_conference}

\begin{thebibliography}{39}
\providecommand{\natexlab}[1]{#1}
\providecommand{\url}[1]{\texttt{#1}}
\expandafter\ifx\csname urlstyle\endcsname\relax
  \providecommand{\doi}[1]{doi: #1}\else
  \providecommand{\doi}{doi: \begingroup \urlstyle{rm}\Url}\fi

\bibitem[Ayed et~al.(2020)Ayed, Stella, Januschowski, and
  Gasthaus]{ayed2020anomaly}
Fadhel Ayed, Lorenzo Stella, Tim Januschowski, and Jan Gasthaus.
\newblock Anomaly detection at scale: The case for deep distributional time
  series models.
\newblock \emph{arXiv preprint arXiv:2007.15541}, 2020.

\bibitem[Bahdanau et~al.(2014)Bahdanau, Cho, and Bengio]{bahdanau2014neural}
Dzmitry Bahdanau, Kyunghyun Cho, and Yoshua Bengio.
\newblock Neural machine translation by jointly learning to align and
  translate.
\newblock \emph{arXiv preprint arXiv:1409.0473}, 2014.

\bibitem[Chandola et~al.(2009)Chandola, Banerjee, and
  Kumar]{chandola2009anomaly}
Varun Chandola, Arindam Banerjee, and Vipin Kumar.
\newblock Anomaly detection: A survey.
\newblock \emph{ACM computing surveys (CSUR)}, 41\penalty0 (3):\penalty0 1--58,
  2009.

\bibitem[DCASE(2020)]{urldcase}
DCASE.
\newblock Task 2 challenge description web page - unsupervised detection of
  anomalous sounds for machine condition monitoring, 2020.
\newblock URL
  \url{http://dcase.community/challenge2020/task-unsupervised-detection-/-of-anomalous-sounds}.

\bibitem[Germain et~al.(2015)Germain, Gregor, Murray, and
  Larochelle]{germain2015made}
Mathieu Germain, Karol Gregor, Iain Murray, and Hugo Larochelle.
\newblock Made: Masked autoencoder for distribution estimation.
\newblock pp.\  881--889, 2015.

\bibitem[Giri et~al.(2020{\natexlab{a}})Giri, Cheng, Helwani, Tenneti, Isik,
  and Krishnaswamy]{Giri2020gmade}
Ritwik Giri, Fangzhou Cheng, Karim Helwani, Srikanth~V. Tenneti, Umut Isik, and
  Arvindh Krishnaswamy.
\newblock Unsupervised anomalous sound detection using self-supervised
  classification and group masked autoencoder based density estimation for
  audio anomaly detection.
\newblock \emph{DCASE2020 Challenge}, July 2020{\natexlab{a}}.

\bibitem[Giri et~al.(2020{\natexlab{b}})Giri, Cheng, Helwani, Tenneti, Isik,
  and Krishnaswamy]{gmade}
Ritwik Giri, Fangzhou Cheng, Karim Helwani, Srikanth~V. Tenneti, Umut Isik, and
  Arvindh Krishnaswamy.
\newblock Group masked auto-encoder based density estimation for audio anomaly
  detection.
\newblock In \emph{sumbitted to DCASE 2020 Workshop}, 2020{\natexlab{b}}.

\bibitem[Giri et~al.(2020{\natexlab{c}})Giri, Tenneti, Helwani, Cheng, Isik,
  and Krishnaswamy]{Giri2020}
Ritwik Giri, Srikanth~V. Tenneti, Karim Helwani, Fangzhou Cheng, Umut Isik, and
  Arvindh Krishnaswamy.
\newblock Unsupervised anomalous sound detection using self-supervised
  classification and group masked autoencoder for density estimation.
\newblock \emph{DCASE2020 Challenge}, July 2020{\natexlab{c}}.

\bibitem[Gong et~al.(2019)Gong, Liu, Le, Saha, Mansour, Venkatesh, and
  Hengel]{gong2019memorizing}
Dong Gong, Lingqiao Liu, Vuong Le, Budhaditya Saha, Moussa~Reda Mansour, Svetha
  Venkatesh, and Anton van~den Hengel.
\newblock Memorizing normality to detect anomaly: Memory-augmented deep
  autoencoder for unsupervised anomaly detection.
\newblock In \emph{Proceedings of the IEEE International Conference on Computer
  Vision}, pp.\  1705--1714, 2019.

\bibitem[Goodge et~al.(2020)Goodge, Hooi, Ng, and Ng]{goodge2020}
Adam Goodge, Bryan Hooi, See~Kiong Ng, and Wee~Siong Ng.
\newblock Robustness of autoencoders for anomaly detection under adversarial
  impact.
\newblock In Christian Bessiere (ed.), \emph{Proceedings of the Twenty-Ninth
  International Joint Conference on Artificial Intelligence, {IJCAI-20}}, pp.\
  1244--1250. International Joint Conferences on Artificial Intelligence
  Organization, 7 2020.
\newblock \doi{10.24963/ijcai.2020/173}.
\newblock URL \url{https://doi.org/10.24963/ijcai.2020/173}.
\newblock Main track.

\bibitem[Gustafsson(2000)]{gustafsson2000adaptive}
Fredrik Gustafsson.
\newblock \emph{Adaptive filtering and change detection}, volume~1.
\newblock Citeseer, 2000.

\bibitem[Hampel(1971)]{hampel1971general}
Frank~R Hampel.
\newblock A general qualitative definition of robustness.
\newblock \emph{The Annals of Mathematical Statistics}, pp.\  1887--1896, 1971.

\bibitem[He et~al.(2015)He, Zhang, Ren, and Sun]{resnet}
Kaiming He, Xiangyu Zhang, Shaoqing Ren, and Jian Sun.
\newblock Deep residual learning for image recognition.
\newblock \emph{CoRR}, abs/1512.03385, 2015.
\newblock URL \url{http://arxiv.org/abs/1512.03385}.

\bibitem[Huber \& Ronchetti(1981)Huber and Ronchetti]{huber1981robust}
Peter~J Huber and E~Ronchetti.
\newblock Robust statistics john wiley \& sons.
\newblock \emph{New York}, 1\penalty0 (1), 1981.

\bibitem[{Jadidi} et~al.(2013){Jadidi}, {Muthukkumarasamy}, {Sithirasenan}, and
  {Sheikhan}]{6679866}
Z.~{Jadidi}, V.~{Muthukkumarasamy}, E.~{Sithirasenan}, and M.~{Sheikhan}.
\newblock Flow-based anomaly detection using neural network optimized with gsa
  algorithm.
\newblock pp.\  76--81, 2013.

\bibitem[Kailath(1968)]{kailath1968innovations}
Thomas Kailath.
\newblock An innovations approach to least-squares estimation--part i: Linear
  filtering in additive white noise.
\newblock \emph{IEEE transactions on automatic control}, 13\penalty0
  (6):\penalty0 646--655, 1968.

\bibitem[Kalchbrenner et~al.(2018)Kalchbrenner, Elsen, Simonyan, Noury,
  Casagrande, Lockhart, Stimberg, Oord, Dieleman, and
  Kavukcuoglu]{kalchbrenner2018efficient}
Nal Kalchbrenner, Erich Elsen, Karen Simonyan, Seb Noury, Norman Casagrande,
  Edward Lockhart, Florian Stimberg, Aaron van~den Oord, Sander Dieleman, and
  Koray Kavukcuoglu.
\newblock Efficient neural audio synthesis.
\newblock \emph{arXiv preprint arXiv:1802.08435}, 2018.

\bibitem[Koizumi et~al.(2019)Koizumi, Saito, Uematsu, Harada, and
  Imoto]{ToyADMOS}
Yuma Koizumi, Shoichiro Saito, Hisashi Uematsu, Noboru Harada, and Keisuke
  Imoto.
\newblock Toyadmos: A dataset of miniature-machine operating sounds for
  anomalous sound detection.
\newblock In \emph{2019 IEEE Workshop on Applications of Signal Processing to
  Audio and Acoustics (WASPAA)}, pp.\  313--317. IEEE, 2019.

\bibitem[Koizumi et~al.(2020)Koizumi, Kawaguchi, Imoto, Nakamura, Nikaido,
  Tanabe, Purohit, Suefusa, Endo, Yasuda, and Harada]{Koizumi_DCASE2020_01}
Yuma Koizumi, Yohei Kawaguchi, Keisuke Imoto, Toshiki Nakamura, Yuki Nikaido,
  Ryo Tanabe, Harsh Purohit, Kaori Suefusa, Takashi Endo, Masahiro Yasuda, and
  Noboru Harada.
\newblock Description and discussion on {DCASE}2020 challenge task2:
  Unsupervised anomalous sound detection for machine condition monitoring.
\newblock In \emph{arXiv e-prints: 2006.05822}, pp.\  1--4, June 2020.

\bibitem[Liu et~al.(2008)Liu, Ting, and Zhou]{liu2008IF}
Fei~Tony Liu, Kai~Ming Ting, and Zhi-Hua Zhou.
\newblock Isolation forest.
\newblock ICDM '08, pp.\  413–422, USA, 2008. IEEE Computer Society.
\newblock ISBN 9780769535029.
\newblock \doi{10.1109/ICDM.2008.17}.
\newblock URL \url{https://doi.org/10.1109/ICDM.2008.17}.

\bibitem[Manevitz \& Yousef(2001)Manevitz and Yousef]{manevitz2001one}
Larry~M Manevitz and Malik Yousef.
\newblock One-class svms for document classification.
\newblock \emph{Journal of machine Learning research}, 2\penalty0
  (Dec):\penalty0 139--154, 2001.

\bibitem[Marchi et~al.(2015)Marchi, Vesperini, Weninger, Eyben, Squartini, and
  Schuller]{marchi2015non}
Erik Marchi, Fabio Vesperini, Felix Weninger, Florian Eyben, Stefano Squartini,
  and Bj{\"o}rn Schuller.
\newblock Non-linear prediction with lstm recurrent neural networks for
  acoustic novelty detection.
\newblock In \emph{2015 International Joint Conference on Neural Networks
  (IJCNN)}, pp.\  1--7. IEEE, 2015.

\bibitem[McFee et~al.(2015)McFee, Raffel, Liang, Ellis, McVicar, Battenberg,
  and Nieto]{mcfee2015librosa}
Brian McFee, Colin Raffel, Dawen Liang, Daniel~PW Ellis, Matt McVicar, Eric
  Battenberg, and Oriol Nieto.
\newblock librosa: Audio and music signal analysis in python.
\newblock In \emph{Proceedings of the 14th python in science conference},
  volume~8, 2015.

\bibitem[Murphy(2012)]{murphy2012machine}
Kevin~P Murphy.
\newblock \emph{Machine learning: a probabilistic perspective}.
\newblock MIT press, 2012.

\bibitem[Nalisnick et~al.(2019)Nalisnick, Matsukawa, Teh, Gorur, and
  Lakshminarayanan]{nalisnick2018deep}
Eric Nalisnick, Akihiro Matsukawa, Yee~Whye Teh, Dilan Gorur, and Balaji
  Lakshminarayanan.
\newblock Do deep generative models know what they don't know?
\newblock In \emph{International Conference on Learning Representations}, 2019.
\newblock URL \url{https://openreview.net/forum?id=H1xwNhCcYm}.

\bibitem[Oliva et~al.(2017)Oliva, Dubey, P{\'o}czos, Xing, and
  Schneider]{oliva2017recurrent}
Junier~B Oliva, Kumar~Avinava Dubey, Barnab{\'a}s P{\'o}czos, Eric Xing, and
  Jeff Schneider.
\newblock Recurrent estimation of distributions.
\newblock \emph{arXiv preprint arXiv:1705.10750}, 2017.

\bibitem[Papamakarios et~al.(2017)Papamakarios, Pavlakou, and
  Murray]{papamakarios2017masked}
George Papamakarios, Theo Pavlakou, and Iain Murray.
\newblock Masked autoregressive flow for density estimation.
\newblock In \emph{Advances in Neural Information Processing Systems}, pp.\
  2338--2347, 2017.

\bibitem[Peel \& McLachlan(2000)Peel and McLachlan]{peel2000robust}
David Peel and Geoffrey~J McLachlan.
\newblock Robust mixture modelling using the t distribution.
\newblock \emph{Statistics and computing}, 10\penalty0 (4):\penalty0 339--348,
  2000.

\bibitem[Purohit et~al.(2019)Purohit, Tanabe, Ichige, Endo, Nikaido, Suefusa,
  and Kawaguchi]{Purohit_DCASE2019_01}
Harsh Purohit, Ryo Tanabe, Takeshi Ichige, Takashi Endo, Yuki Nikaido, Kaori
  Suefusa, and Yohei Kawaguchi.
\newblock {MIMII Dataset}: Sound dataset for malfunctioning industrial machine
  investigation and inspection.
\newblock In \emph{({DCASE2019})}, pp.\  209--213, November 2019.

\bibitem[Rezende \& Mohamed(2015)Rezende and Mohamed]{rezende2015variational}
Danilo~Jimenez Rezende and Shakir Mohamed.
\newblock Variational inference with normalizing flows.
\newblock \emph{arXiv preprint arXiv:1505.05770}, 2015.

\bibitem[Salinas et~al.(2020)Salinas, Flunkert, Gasthaus, and
  Januschowski]{salinas2020deepar}
David Salinas, Valentin Flunkert, Jan Gasthaus, and Tim Januschowski.
\newblock Deepar: Probabilistic forecasting with autoregressive recurrent
  networks.
\newblock \emph{International Journal of Forecasting}, 36\penalty0
  (3):\penalty0 1181--1191, 2020.

\bibitem[Sandler et~al.(2018)Sandler, Howard, Zhu, Zhmoginov, and Chen]{mobile}
Mark Sandler, Andrew~G. Howard, Menglong Zhu, Andrey Zhmoginov, and
  Liang{-}Chieh Chen.
\newblock Inverted residuals and linear bottlenecks: Mobile networks for
  classification, detection and segmentation.
\newblock \emph{CoRR}, abs/1801.04381, 2018.
\newblock URL \url{http://arxiv.org/abs/1801.04381}.

\bibitem[Svens{\'e}n \& Bishop(2005)Svens{\'e}n and Bishop]{svensen2005robust}
Markus Svens{\'e}n and Christopher~M Bishop.
\newblock Robust bayesian mixture modelling.
\newblock \emph{Neurocomputing}, 64:\penalty0 235--252, 2005.

\bibitem[Wong et~al.(2009)Wong, Chan, and Kam]{wong2009student}
CS~Wong, WS~Chan, and PL~Kam.
\newblock A student t-mixture autoregressive model with applications to
  heavy-tailed financial data.
\newblock \emph{Biometrika}, 96\penalty0 (3):\penalty0 751--760, 2009.

\bibitem[Wu et~al.(2019)Wu, Gao, Han, and Zhu]{wu2019tale}
Ying~Nian Wu, Ruiqi Gao, Tian Han, and Song-Chun Zhu.
\newblock A tale of three probabilistic families: Discriminative, descriptive,
  and generative models.
\newblock \emph{Quarterly of Applied Mathematics}, 77\penalty0 (2):\penalty0
  423--465, 2019.

\bibitem[Yamaguchi et~al.(2019)Yamaguchi, Koizumi, and
  Harada]{yamaguchi2019adaflow}
Masataka Yamaguchi, Yuma Koizumi, and Noboru Harada.
\newblock Adaflow: Domain-adaptive density estimator with application to
  anomaly detection and unpaired cross-domain translation.
\newblock pp.\  3647--3651. IEEE, 2019.

\bibitem[Zhang et~al.(2019)Zhang, Song, Chen, Feng, Lumezanu, Cheng, Ni, Zong,
  Chen, and Chawla]{zhang2019deep}
Chuxu Zhang, Dongjin Song, Yuncong Chen, Xinyang Feng, Cristian Lumezanu, Wei
  Cheng, Jingchao Ni, Bo~Zong, Haifeng Chen, and Nitesh~V Chawla.
\newblock A deep neural network for unsupervised anomaly detection and
  diagnosis in multivariate time series data.
\newblock In \emph{Proceedings of the AAAI Conference on Artificial
  Intelligence}, volume~33, pp.\  1409--1416, 2019.

\bibitem[Zhou \& Paffenroth(2017)Zhou and Paffenroth]{zhou2017anomaly}
Chong Zhou and Randy~C Paffenroth.
\newblock Anomaly detection with robust deep autoencoders.
\newblock In \emph{Proceedings of the 23rd ACM SIGKDD International Conference
  on Knowledge Discovery and Data Mining}, pp.\  665--674, 2017.

\bibitem[Zong et~al.(2018)Zong, Song, Min, Cheng, Lumezanu, Cho, and
  Chen]{zong2018deep}
Bo~Zong, Qi~Song, Martin~Renqiang Min, Wei Cheng, Cristian Lumezanu, Daeki Cho,
  and Haifeng Chen.
\newblock Deep autoencoding gaussian mixture model for unsupervised anomaly
  detection.
\newblock In \emph{International Conference on Learning Representations}, 2018.

\end{thebibliography}
\bibliographystyle{iclr2021_conference}
\clearpage
\appendix

\section{Temporal Attention Mechanism}\label{TAM}
To help the network finding relevant pattern in long sequential data, we augment the recurrent layers with the attention mechanism presented in \cite{bahdanau2014neural}. In this mechanism, while processing a sequence of length $L$ the hidden states of the RNN, $\mathbf{h}_t$, are augmented with normalized weights a.k.a. attention vector $\boldsymbol{\beta}=[{\beta}_1,\ldots,{\beta}_L]$ which is calculated as follows%= [\mathbf{h}_{t,\ldots,t-L+1}]%
\begin{align}
\begin{array}{c}
\mathbf{q}_l=\tanh \left(\mathbf{W}_{a}^l  \mathbf{h}_t\right) \text {, and } {\beta}_l=\operatorname{softmax}\left(\mathbf{W}_{q}^l \mathbf{q}_l\right), 
\end{array}
\end{align}
where $\mathbf{W}_{a}^l$  and $\mathbf{W}_{q}^l$  are learnable weights,  $ \mathbf{q}_l$  is a learned embedding for each of the hidden states, and $\operatorname{tanh}$ is a hyperbolic tangent activation function. Finally, this attention vector is applied to the states in a sequence as following to obtain a new state vector as an average of the hidden states weighted by $\boldsymbol{\beta}$ or attention vector
\begin{align}
\mathbf{h}=\sum_{l=1}^{L} {\beta}_l\mathbf{h}_l.
\end{align}

\section{Multiresolutional Architecture} \label{MA}
\begin{figure}[h]
	\centering
	\includegraphics[width=0.4\columnwidth]{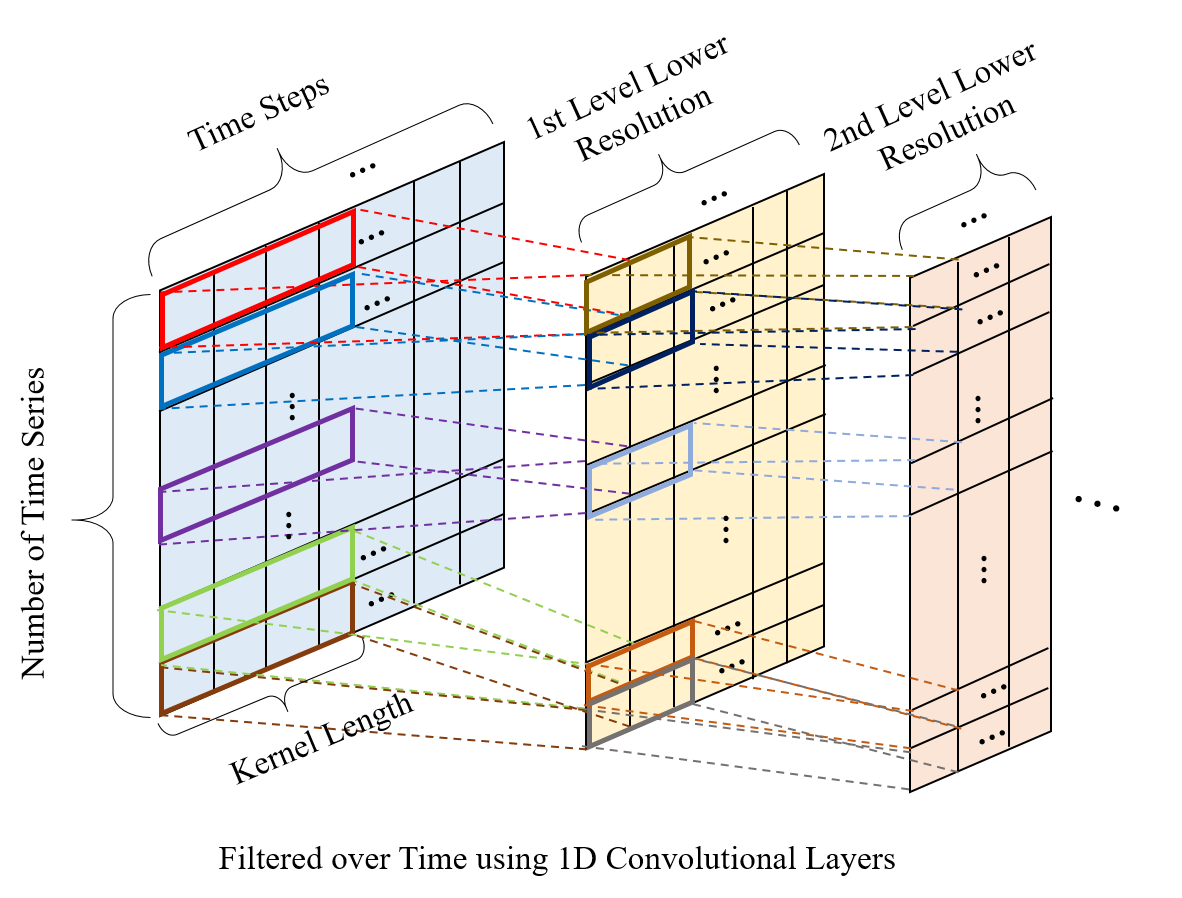}
	\caption{Multiresolution features extraction over time}
	\label{fig2}
\end{figure}
The network we propose in this paper is built in a hierarchical manner allowing the extraction of global features (Multiresolution). We achieve this by replicating the original network and feeding it with the inputs after passing them through a temporal filter. As shown in Fig. ~\ref{fig2}, 1D filters are applied to each time series to extract features over time. The extracted features are fed into GRU as the original network does. The final output layer is then conditioned on the output of these additional networks. 

\section{Datasets, data preprocessing, and evaluation metrics}\label{dataset}
The dataset is organized in sound recordings where each recording is a single-channel 10-sec length audio which contains both a target machine's operating sound and environmental noise. The following six types of toy/real machines were used; Toy-car (\texttt{ToyADMOS}), Toy-conveyor (\texttt{ToyADMOS}), Valve (\texttt{MIMII}), Pump (\texttt{MIMII}), Fan (\texttt{MIMII}), Slide rail (\texttt{MIMII}) \citep{urldcase,Koizumi_DCASE2020_01,ToyADMOS,Purohit_DCASE2019_01}. There are three machine IDs in each machine type, so the total number of datasets used for training and evaluation is 18. The given task is to detect unknown anomalous sounds that are not observed during the training phase. This can be used to determine whether a condition of target machine is normal or anomalous (e.g., machine failure). For training, only recordings from machines operating under normal conditions are used. More detailed descriptions of the datasets (\texttt{ToyADMOS} and \texttt{MIMII}) and data reprocessing are as follow.

\begin{itemize}[leftmargin=0.3cm, topsep=-1pt]
\item	\texttt{ToyADMOS} \citep{ToyADMOS}. The dataset is provided for audio based machine anomaly detection. The dataset consists of normal operating sounds and anomalous operating sounds. For training, only normal operating sounds are  used. The anomalous machine operation sounds were collected by deliberately introducing a fault (e.g., bent shaft, deformation of gear, and excessive tension) in a machine. To simulate actual factory environment, the sounds were randomly mixed with various kinds of environmental noise, which were collected from several places in a real factory. Multivariate time-series are extracted from the raw audio signal either as log Mel or MFCC features using the librosa python library package \citep{mcfee2015librosa}; log-Mel feature is used for Fan and Pump and MFCC feature is used for slider and valve. The number of Mel and MFCC is set to 90 (i.e., 90 dimensions), and the features are scaled to a range between -1 and 1.

\item	\texttt{MIMII} \citep{Purohit_DCASE2019_01}. MIMII is the industrial machine sounds dataset that contains  sounds of industrial equipment (e.g., valves, pumps, fans, and sliders). The sound of industrial equipment is recorded under normal and anomalous operating conditions. The malfunctions causing the anomalous sound included mechanical unbalanced and looseness in the industrial equipment. Various background noises, which were collected from multiple locations in a factory, were mixed with the recorded sounds to simulate a variety of noisy conditions. We extracted log Mel features from the raw audio signal, and the same preprocessing and scaling methods are used as in \texttt{ToyADMOS}.
\end{itemize}

For evaluation metrics, we use the same metrics as proposed by the organizing committee of the DCASE challenge \citep{urldcase}, which are Area Under the ROC Curve (AUC) and partial AUC (pAUC). The AUC and pAUC are defined as follows 
\begin{align}
\begin{array}{c}
\mathrm{AUC}=\frac{1}{N_{-} N_{+}} \sum_{i=1}^{N_{-}} \sum_{j=1}^{N_{+}} \mathcal{H}\left(\mathcal{A}_{\theta}\left(x_{j}^{+}\right)-\mathcal{A}_{\theta}\left(x_{i}^{-}\right)\right) \\
\mathrm{pAUC}=\frac{1}{\left\lfloor p N_{-}\right\rfloor N_{+}} \sum_{i=1}^{{\left\lfloor p N_{-}\right\rfloor }} \sum_{j=1}^{N_{+}} \mathcal{H}\left(\mathcal{A}_{\theta}\left(x_{j}^{+}\right)-\mathcal{A}_{\theta}\left(x_{i}^{-}\right)\right)\nonumber
\end{array}
\end{align}
where $\lfloor\cdot\rfloor$ is the flooring function and $\mathcal{H}(x)$ returns 1 when $x > 0$ and $0$ otherwise, and $\mathcal{A}_{\theta}$ denotes the model output given the parameters $\theta$. Here, $\left\{x_{i}^{-}\right\}_{i=1}^{N_{-}}$ and $\left\{x_{j}^{+}\right\}_{j=1}^{N_{+}}$ are normal and anomalous test samples, respectively, and have been sorted so that their anomaly scores are in descending order. Here, $N_{-}$ and $N_{+}$ are the number of normal and anomalous test samples, respectively. According to the above formulas, the anomaly scores of normal test samples are used as the threshold for a final decision whether a sample is anomalous or normal.

\section{Baseline models}\label{base}
\textbf{OC-SVM}. One-class support vector machine (OC-SVM) \citep{manevitz2001one} is an unsupervised algorithm learning a decision boundary for anomaly/novelty detection. In this study, we test several kernel types (e.g., linear, RBF, and polynomial), and we simply choose a kernel that renders the best performance. In our experiment, OC-SVM with RBF kernel shows the best performance.       

\textbf{IF}. Isolation Forest (IF) \citep{liu2008IF} is a unsupervised anomaly detection algorithm, which explicitly isolates anomalies in the dataset using by means of decision trees. In the method, an ensemble of isolation trees is used to isolate anomalies. To train the model, several parameters (e.g., contamination, the number of trees in ensemble, and maximum number of samples to be drawn to train each base estimator) need to be set, so we simply conduct exhaustive search to find the optimal parameters, and following parameters are adopted: the number of trees = 150, max sample = 256, contamination = 0.05).

\textbf{AE}. Autoencoder (AE) learns a representation of the data, and it's trained to reconstruct an original input. After training, an error between  input and output can used as an anomaly score. The method is based on the assumption that anomaly data cannot be effectively reconstructed. We used the same number of layers as done in the DCASE baseline in our AE model, and it is FC(90, 90, relu)-FC(90, 60, relu)-FC(60, 30, relu)-FC(30, 15, relu)-FC(15, 30, relu)-FC(30, 60, relu)-FC(60, 90, relu)-FC(90,90, none), where FC is a fully connected layer. In the training phase, following parameters are used: learning rate = 0.001, optimizer = adam, batch size = 64, regularization = L2, epoch = 50, and loss function = mean squared error (L2). 

\textbf{GMM}. Gaussian Mixture Model (GMM) is a density estimation model to model a complex distribution using multiple Gaussian distributions. Two key parameters of GMM are the number of mixture components and covariance type (full, tied, diagonal, and spherical). We simply perform exhaustive search to find optimal parameters, and ten components with full covariance type are adopted in our study. An expectation–maximization (EM) algorithm is used to fit the model using the training dataset, and log-likelihood is used as an anomaly score during the inference.

\textbf{MSCRED}. \citep{zhang2019deep} Multi-Scale Convolutional Recurrent Encoder-Decoder (MSCRED) reconstructs multi-scale signature matrices, which encode the inter-variable correlations and the temporal dependency. A reconstruction error is used as an anomaly score. The publicly available code is used to implement the model. The size of signature matrices is modified according to the number of dimensions we analyzed. For training parameters, we borrow the wisdom from the original implementation.   

\textbf{DAGMM}. \citep{zong2018deep} Deep Autoencoding Gaussian Mixture Model (DAGMM) is a density based model. Sample energy is used as an anomaly score. The publicly available code is used to implement the model. For training parameters, we borrow the wisdom from the original implementation.  

\textbf{DCASE winner}. \citep{Giri2020} The system is an ensemble of self-supervised models using MobileNetV2, ResnetV2, and a multivariate density estimator GroupMADE. The model is trained for an auxiliary classification task which fits into the self-supervision  framework. During the model training, several data augmentation techniques such as synthetic augmentation techniques and spectral warping augmentation were considered, more details can be found in the cited paper. 

\textbf{DCASE baseline}. \citep{Koizumi_DCASE2020_01} The DCASE baseline is an autoencoder based model, which uses sample reconstruction error of the observed sound as an anomaly score. Details on the model can be found in the cited paper.

\section{Implementation details}\label{implementation}
The architecture shown in Fig. \ref{fig1} presents an overview of the proposed model. We implement the proposed model for the task of anomaly detection. We describe the implementation in four parts; (a) estimating the temporal dynamic of the time series, (b) multiresolutional feature extraction, (c) parameter estimation for a mixture of multivariate distributions, and (d) NLL loss calculation. 

In (a), the multivariate time series, $\mathbf{x}_{t-1:0}$, are passed through the attention augmented recurrent layers. The recurrent architecture used in the model is based on gated recurrent unit (GRU), and the number of hidden layers, sequence length, and hidden size are set to 2, 70 (i.e., 70 frames), and 512. The attention vector is calculated as described in Appendix \ref{TAM}. 

(b) is a parallel process with (a). In (b), 1D-Convolutional layers (Conv) with kernel length = 10, stride = 3, and padding = 0 are used to extract lower resolution features as described in Appendix \ref{MA}. Each frequency bin (i.e., dimension of the time series is treated as univariate time series) is processed with a 1-D convolutional layer; as shown in Fig. ~\ref{fig2}, 1D filters are applied to each time series to extract features over time. The output of the convolutional layer is passed through GRU layers similar to (a). As in (a), the outputs of the GRU layer are augmented with an attention mechanism. Outputs from (a) and (b) are concatenated and fed into a fully connected layer for the pdf parameters estimation. 

In (c), the parameters of a mixture of multivariate Student-$t$ distribution, which are means vector, covariance matrices, degrees of freedom, and responsibilities, are estimated as described in Section \ref{model_archi}. While the means vector and the lower triangular part of covariance matrices are simply estimated using fully connected layers, the specialized activation functions are additionally considered to impose constraints on the diagonal components of covariance matrices, the degrees of freedom, and the responsibilities. For example, the softplus activation function is placed after the fully connected layer to output only positive values for the diagonal components of the covariance matrix. Also, the scaled sigmoid activation function is used to set upper and lower bounds of the degree of freedom (or, to avoid an extremely large degrees of freedom, which would make Student $t$ distribution similar to a Gaussian distribution). In this study, the lower and upper bounds of the degree of freedom are set between 1 and 10. The softmax activation is used to calculate the responsibilities of each mixture components, hence, these values are all positive and sum to 1. The output size of the network for each parameter is a function of the number of components in a mixture ($c$) and the dimensionality of the data ($P$). In our experiments, $c$ is set to 3.   

In (d), with the parameters estimated by the model in (c), a mixture model is obtained using Eq. (\ref{eq:9}), and an NLL loss is calculated for an observed value, $\boldsymbol{x}_t$, through Eq. (\ref{eq:12}).

The proposed model (RSMM-MR) and the four variants (RGMM, RGMM-MR, RSMM, RSMM-MR-W/o Attention) are trained with the training parameters of epoch = 30, batch size = 128, initialization = uniform with the magnitude of 0.1, optimizer = Adam, and learning rate = 0.00001. L2 regularization with weight decay parameter = 0.001 is chosen. The training and inference are done on a Tesla V100 GPU.

In the implementation of the ensemble of our model with the DCASE winner model \citep{Giri2020}, we transform the anomaly scores of each model in the ensemble into a standardized scale, before combining them. The standardization transformation for any given model is applied in a per-machine manner where we calculate the mean and variance of its anomaly scores over the training data for that machine ID. The anomaly scores are then transformed to have zero mean and unit variance over the training data of that machine ID. Standardized anomaly scores across different models are then combined using mean or max ensembling.

\section{Detailed results of anomaly detection experiment}\label{allexp}
For the anomaly detection task, we use the machine sounds datasets (\texttt{ToyADMOS} and \texttt{MIMII}). We present the average performance of each model in Table \ref{table1_all}. Our proposed model, RSMM-MR, demonstrates the superior performance over the most of the baseline models, and shows the comparable results to the DCASE winner. The DCASE winner is an ensemble of MobileNetV2 \citep{mobile}, ResnetV2 \citep{resnet}, and GroupMADE \citep{gmade}, and also most top-ranked models in the challenge are ensembles of different anomaly detector. Among the traditional anomaly detection models, GMM, which is a density based model, performs the best in all settings in terms of AUC / pAUC, followed by AE, IF, and OC-SVM. While the GMM works well in many datasets, RSMM-MR outperforms in all datasets, especially improving AUC / pAUC by 0.2286 / 0.3417 and 0.4074 / 0.3679 in Slider and Valve, respectively. In our experiment, our approach outperforms MSCRED and DAGMM. %The possible reasons could be an existence of random environmental noise in both training and testing data. 
In \citet{zong2018deep}, DAGMM was able to reduce the effect of contamination in trading data, but the anomaly detection accuracy gradually decreased with increasing contamination ratio. DAGMM  performed similarly in benchmarks conducted in two recent anomaly detection papers \citep{goodge2020,zhang2019deep}. Also, ensembling the proposed model with the
DCASE winner model further improves the results in all cases.

In Table \ref{table3}, we report the results on each machine IDs (there are three machine IDs in each machine type). Results of the DCASE winner on each machine IDs were not reported by the authors, hence, we omit this model in Table \ref{table3}. As it can be observed, our proposed model outperforms the baseline models consistently in most cases. %Because the anomalies in each ID were induced by the different causes, this result indicates that the proposed model effectively detect various types / statistics of anomalies. 
Confirming that the model learns well the representation of the noisy training data (i.e., normal data) without either generalizing to unseen anomalies data nor overfitting on the training dataset.

\begin{table*}[t]
\caption{Performance of the proposed system on machine IDs measured in AUC / pAUC and comparison to the baseline models (bold numbers are the highest number on each ID)}
\begin{center}
\begin{small}
\begin{sc}
\
\resizebox{1\textwidth}{!}{%
\begin{tabular}{|c||c|c|c|c|c||c|c||c|c|}

\hline
\multirow{2}{*}{\textbf{Method}}                & \multirow{2}{*}{\textbf{ID}} & \multicolumn{4}{c||}{\textbf{MIMII}}                                            & \multirow{2}{*}{\textbf{ID}} & \textbf{TOYADMOS}        & \multirow{2}{*}{\textbf{ID}} & \textbf{TOYADMOS}        \\ \cline{3-6} \cline{8-8} \cline{10-10} 
                                       &                     & \textbf{FAN}             & \textbf{PUMP}            & \textbf{SLIDER}          & \textbf{VALVE}           &                     & \textbf{TOY CONVEYOR}    &                     & \textbf{TOY CAR}         \\ \hline
\multirow{3}{*}{OC\_SVM}               & 1                   & 0.6087 / 0.5634 & 0.7824 / 0.6529 & 0.5969 / 0.5115 & 0.3957 / 0.4781 & 4                   & 0.5192 / 0.4914 & 5                   & 0.5894 / 0.5295 \\ \cline{2-10} 
                                       & 3                   & 0.6589 / 0.6253 & 0.6398 / 0.5021 & 0.5742 / 0.4920 & 0.4724 / 0.5132 & 5                   & 0.5756 / 0.5307 & 6                   & 0.6478 / 0.5318 \\ \cline{2-10} 
                                       & 5                   & 0.6688 / 0.5584 & 0.5094 / 0.5488 & 0.5597 / 0.5050 & 0.5147 / 0.4997 & 6                   & 0.4937 / 0.5040 & 7                   & 0.6195 / 0.5341 \\ \hline
\multirow{3}{*}{IF}                    & 1                   & 0.6512 / 0.5032 & 0.7680 / 0.6284 & 0.6178 / 0.5044 & 0.4125 / 0.4886 & 4                   & 0.6090 / 0.5188 & 5                   & 0.6268 / 0.5257 \\ \cline{2-10} 
                                       & 3                   & 0.7220 / 0.5638 & 0.7028 / 0.4886 & 0.5825 / 0.4979 & 0.4812 / 0.5018 & 5                   & 0.6138 / 0.5121 & 6                   & 0.7232 / 0.5461 \\ \cline{2-10} 
                                       & 5                   & 0.6716 / 0.5453 & 0.6171 / 0.5700 & 0.5327 / 0.5062 & 0.4984 / 0.4951 & 6                   & 0.5230 / 0.5066 & 7                   & 0.7324 / 0.5568 \\ \hline
\multirow{3}{*}{AE}                    & 1                   & 0.7167 / 0.5644 & 0.8147 / 0.6779 & 0.7129 / 0.5452 & 0.5391 / 0.5193 & 4                   & 0.8832 / 0.7437 & 5                   & 0.6744 / 0.5610 \\ \cline{2-10} 
                                       & 3                   & 0.7857 / 0.6263 & 0.7782 / 0.5813 & 0.6399 / 0.5169 & 0.5273 / 0.5127 & 5                   & 0.8031 / 0.6000 & 6                   & 0.8564 / 0.6513 \\ \cline{2-10} 
                                       & 5                   & 0.8611 / 0.7491 & 0.7911 / 0.6250 & 0.5596 / 0.4950 & 0.5063 / 0.4882 & 6                   & 0.7335 / 0.5860 & 7                   & 0.8556 / 0.6489 \\ \hline
\multirow{3}{*}{GMM}                   & 1                   & 0.7524 / 0.5789 & 0.8868 / 0.7350 & 0.8519 / 0.6446 & 0.5673 / 0.5351 & 4                   & 0.9088 / 0.7900 & 5                   & 0.6873 / 0.5674 \\ \cline{2-10} 
                                       & 3                   & 0.8391 / 0.6498 & 0.8401 / 0.6297 & 0.7153 / 0.5293 & 0.6283 / 0.5298 & 5                   & 0.8048 / 0.6141 & 6                   & 0.8395 / 0.6255 \\ \cline{2-10} 
                                       & 5                   & 0.8857 / \textbf{0.7869} & 0.8783 / 0.6475 & 0.6926 / 0.5166 & 0.5077 / 0.4934 & 6                   & 0.7593 / 0.5868 & 7                   & 0.8784 / 0.6833 \\ \hline
\multirow{3}{*}{MSCRED}                & 1                   & 0.5840 / 0.5260 & 0.5598 / 0.5544 & 0.4205 / 0.4867 & 0.4433 / 0.4750 & 4                   & 0.4631 / 0.4761 & 5                   & 0.4424 / 0.5001 \\ \cline{2-10} 
                                       & 3                   & 0.6159 / 0.5797 & 0.4461 / 0.5147 & 0.5556 / 0.4894 & 0.4673 / 0.5039 & 5                   & 0.4905 / 0.4888 & 6                   & 0.4160 / 0.4993 \\ \cline{2-10} 
                                       & 5                   & 0.4728 / 0.5045 & 0.4760 / 0.4752 & 0.4328 / 0.4944 & 0.5811 / 0.5163 & 6                   & 0.4834 / 0.4885 & 7                   & 0.4328 / 0.4959 \\ \hline
\multirow{3}{*}{DAGMM}                 & 1                   & 0.5994 / 0.5111 & 0.7379 / 0.6252 & 0.8142 / 0.6212 & 0.4280 / 0.5009 & 4                   & 0.5891 / 0.5143 & 5                   & 0.5719 / 0.5280 \\ \cline{2-10} 
                                       & 3                   & 0.6406 / 0.5112 & 0.5858 / 0.5002 & 0.6394 / 0.5293 & 0.4964 / 0.5153 & 5                   & 0.6045 / 0.5187 & 6                   & 0.7111 / 0.5574 \\ \cline{2-10} 
                                       & 5                   & 0.6913 / 0.5870 & 0.7750 / 0.6466 & 0.4941 / 0.4914 & 0.4942 / 0.4900 & 6                   & 0.6733 / 0.5212 & 7                   & 0.7359 / 0.6076 \\ \hline
\multirow{3}{*}{\makecell{DCASE \\ Baseline}}                 & 1                   & 0.7720 / 0.6127 & 0.8526 / 0.6953 & 0.9044 / 0.7202 & 0.5933 / 0.5175 & 4                   & 0.9208 / 0.7226 & 5                   & 0.7426 / 0.6316 \\ \cline{2-10} 
                                       & 3                   & 0.8561 / 0.6650 & 0.7944 / 0.6060 & 0.8196 / 0.5475 & 0.5657 / 0.5152 & 5                   & 0.8335 / 0.6144 & 6                   & 0.8338 / 0.6992 \\ \cline{2-10} 
                                       & 5                   & 0.8560 / 0.6961 & 0.8242 / 0.6220 & 0.6584 / 0.5040 & 0.5622 / 0.4911 & 6                   & 0.8066 / 0.6115 & 7                   & 0.8279 / 0.6543 \\ \hline                                       
                                       
\multirow{3}{*}{RGMM}                  & 1                   & 0.7813 / 0.5589 & 0.8722 / 0.7146 & 0.9392 / 0.8096 & 0.9978 / 0.9882 & 4                   & 0.9197 / 0.7836 & 5                   & 0.7518 / 0.6400 \\ \cline{2-10} 
                                       & 3                   & 0.8379 / 0.6560 & 0.8089 / 0.6060 & 0.9050 / 0.5769 & 0.8932 / 0.6605 & 5                   & 0.7978 / 0.5889 & 6                   & 0.9347 / 0.7914 \\ \cline{2-10} 
                                       & 5                   & 0.8425 / 0.7362 & 0.8554 / 0.6564 & 0.9079 / 0.6863 & 0.7536 / 0.5301 & 6                   & 0.7409 / 0.5761 & 7                   & 0.9123 / 0.7518 \\ \hline
\multirow{3}{*}{RGMM-MR}               & 1                   & 0.7721 / 0.5547 & 0.9011 / 0.7291 & 0.9571 / 0.8415 & 0.9998 / 0.9990 & 4                   & 0.9245 / 0.7792 & 5                   & 0.7685 / 0.6315 \\ \cline{2-10} 
                                       & 3                   & 0.8334 / 0.6376 & 0.8301 / 0.6134 & 0.9169 / 0.6443 & 0.9554 / 0.8425 & 5                   & 0.8595 / 0.6139 & 6                   & 0.9348 / 0.7869 \\ \cline{2-10} 
                                       & 5                   & 0.8738 / 0.7513 & 0.8794 / 0.6511 & 0.9370 / 0.7429 & 0.9165 / 0.7428 & 6                   & 0.8141 / 0.6374 & 7                   & 0.9402 / 0.8048 \\ \hline
\multirow{3}{*}{RSMM}                  & 1                   & 0.7872 / 0.5659 & 0.8795 / 0.7064 & 0.9613 / 0.8746 & 0.9988 / 0.9934 & 4                   & 0.9361 / 0.7954 & 5                   & 0.7682 / 0.6459 \\ \cline{2-10} 
                                       & 3                   & 0.8755 / 0.6898 & 0.8011 / 0.6190 & 0.9573 / 0.8115 & 0.8659 / 0.6123 & 5                   & 0.8799 / 0.6532 & 6                   & 0.9318 / 0.7866 \\ \cline{2-10} 
                                       & 5                   & \textbf{0.9053} / 0.7856 & 0.8909 / 0.6439 & 0.8888 / 0.6502 & 0.7370 / 0.5067 & 6                   & 0.7978 / 0.6009 & 7                   & 0.9414 / 0.8177 \\ \hline
\multirow{3}{*}{\makecell{RSMM-MR- \\ W/o Attention}} & 1                   & 0.8294 / 0.6320 & \textbf{0.9420} / \textbf{0.8412} & 0.9881 / 0.9520 & 0.9997 / 0.9983 & 4                   & 0.9185 / 0.7517 & 5                   & 0.7732 / 0.6630 \\ \cline{2-10} 
                                       & 3                   & 0.9044 / \textbf{0.7456} & 0.8632 / 0.6460 & 0.9718 / 0.8545 & 0.9505 / 0.7798 & 5                   & 0.9148 / 0.7193 & 6                   & 0.9361 / 0.7991 \\ \cline{2-10} 
                                       & 5                   & 0.9012 / 0.7264 & 0.8711 / 0.6373 & 0.9534 / 0.7572 & 0.9090 / 0.6387 & 6                   & 0.8775 / 0.6793 & 7                   & 0.9216 / 0.7934 \\ \hline
\multirow{3}{*}{\textit{RSMM-MR}}               & 1                   & \textbf{0.8387} / \textbf{0.6418} & 0.9319 / 0.8221 & \textbf{0.9928} / \textbf{0.9619} & \textbf{0.9998} / \textbf{0.9991} & 4                   & \textbf{0.9595} / \textbf{0.8630} & 5                   & \textbf{0.8336} / \textbf{0.6955} \\ \cline{2-10} 
                                       & 3                   & \textbf{0.9090} / 0.7378 & \textbf{0.8810} / \textbf{0.6814} & \textbf{0.9824} / \textbf{0.9092} & \textbf{0.9703} / \textbf{0.8640} & 5                   & \textbf{0.9337} / \textbf{0.7520} & 6                   & \textbf{0.9659} / \textbf{0.8774} \\ \cline{2-10} 
                                       & 5                   & 0.8913 / 0.7365 & \textbf{0.9160} / \textbf{0.7025} & \textbf{0.9704} / \textbf{0.8444} & \textbf{0.9555} / \textbf{0.7988} & 6                   & \textbf{0.9017} / \textbf{0.7477} & 7                   & \textbf{0.9598} / \textbf{0.8608} \\ 
\bottomrule
\end{tabular}
\label{table3}}

\end{sc}
\end{small}
\end{center}
\end{table*}

\section{Ablation analysis}\label{abal}
\begin{figure}[h]
	\centering
	\includegraphics[width=0.9\columnwidth]{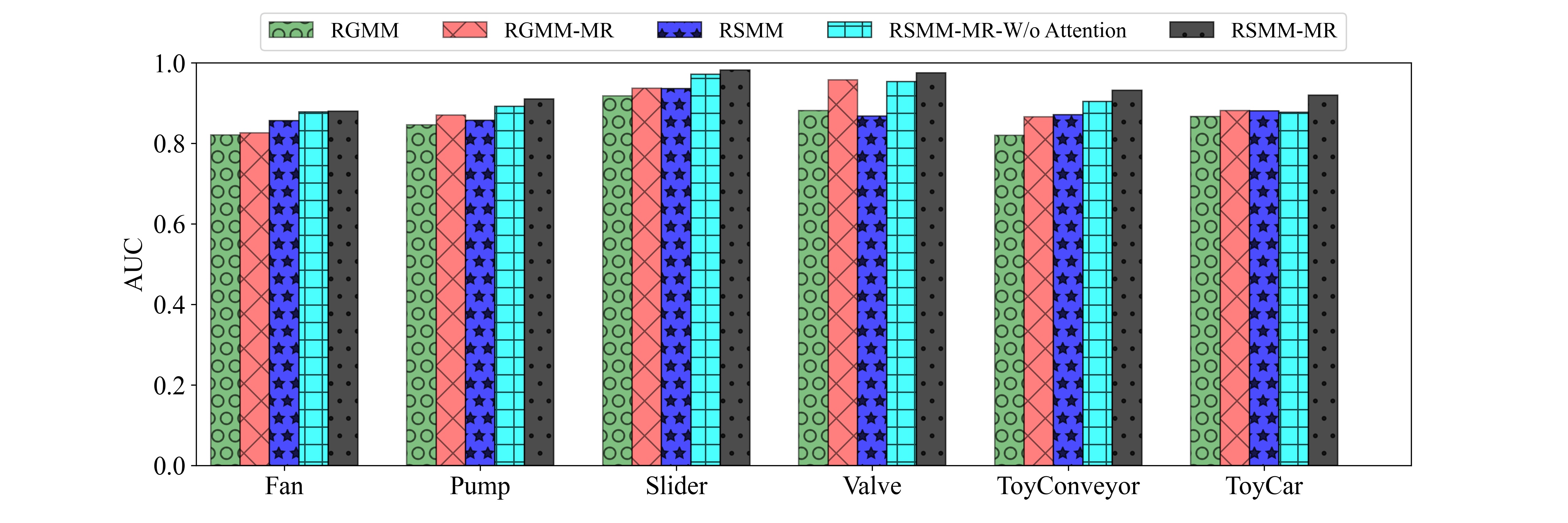} % Reduce the figure size so that it is slightly narrower than the column.
	\caption{Ablation analysis-AUC.}
	\label{fig5}
\end{figure}
\begin{figure}[h]
	\centering
	\includegraphics[width=0.9\columnwidth]{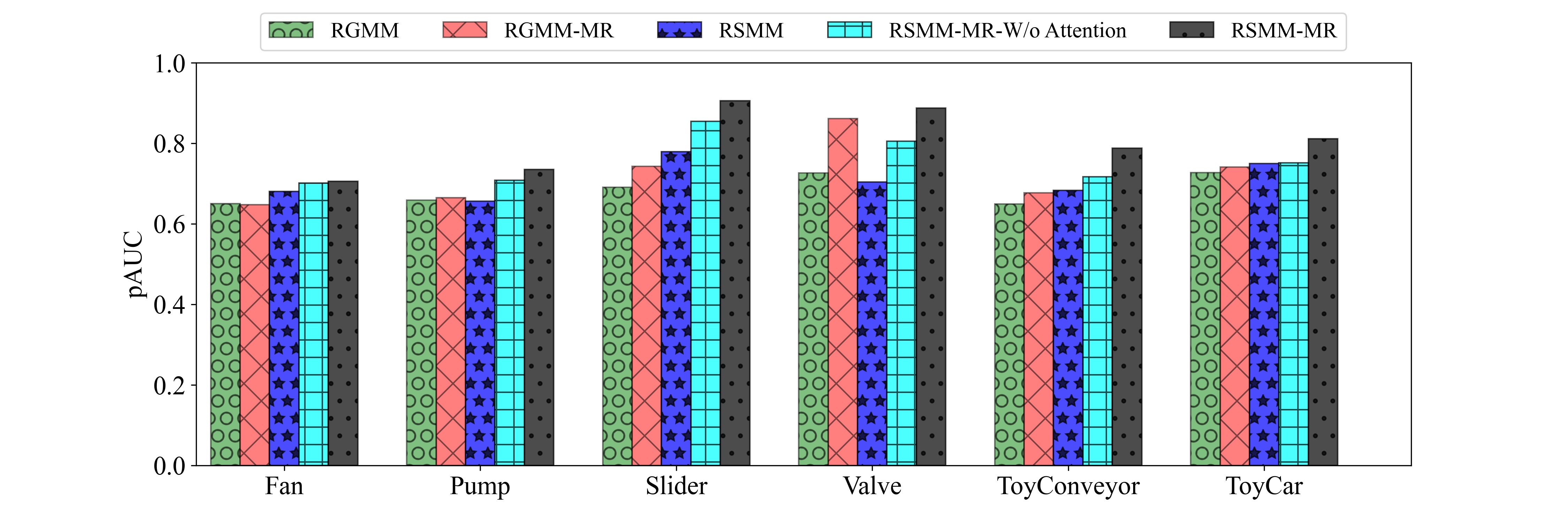} % Reduce the figure size so that it is slightly narrower than the column.
	\caption{Ablation analysis-pAUC.}
	\label{fig6}
\end{figure}

We conduct ablation analysis with respect to the different main features of our model to quantify the influence on the overall achieved model. To this end, four variations of the model are implemented, these are; RGMM (Recurrent GMM - standing for the model presented in this paper using a non-robust GMM model and without using CNN layers for the multiresolutional feature extraction), RGMM-MR (Recurrent GMM with multiresolutional feature extraction - standing for the model presented in this paper using a non-robust GMM model and using CNN layers for multiresolutional feature extraction), RSMM (recurrent Student-t mixture model), and RSMM-MR-W/o Attention. The performance metrics, AUC and pAUC, on each machine types are visualized in Figs. \ref{fig5} and \ref{fig6}.

\subsection{Effect of choosing a robust probability density distribution}
To isolate the effect of choosing an outlier robust pdf in our modelling to the innovation process of the time series the RSMM and RSMM-MR, in which the multivariate Student $t$ mixture model is used, are compared with RGMM and RGMM-MR, in which the multivariate Gaussian mixture model are employed. 

With the same number of mixture components modelling the innovation process, the average of AUC and pAUC observed in RSMM is higher in all cases than those observed in RGMM, except for Valve. The biggest improvement, 0.0184 (AUC) / 0.0879 (pAUC), is observed in Slider. When incorporating the multiresolutional architecture in the models, i.e., RSMM-MR and RGMM-MR, RSMM-MR shows the better performance than RGMM-MR in all cases. Especially, we observe a big improvement in pAUC; the average pAUC of all machine types is higher around 0.0839 in RSMM-MR than RGMM-MR. This empirical results demonstrate the effectiveness of the Student $t$ mixture model in modelling robust parametric distribution.

To further highlight the effectiveness of the choice of outliers robust distribution to deal with noisy training data, we conduct an additional experiment on ToyCar (\texttt{ToyADMOS}),  Slider (\texttt{MIMII}), and Valve (\texttt{MIMII}) where we synthetically introduce noise to the training data. The results are discussed in Appendix \ref{robustness}. 

\subsection{Effect of a multiresolutional architecture}
We also consider the model with and without multiresolutional architecture described in Section \ref{TAM} to demonstrate the effectiveness of extracting global features in the anomaly detection task. Enabling multiresolutional architecture in the model makes the final output conditioned on the output of multiresolutional time-series features.
In all cases, the comparisons between RGMM and RGMM-MR or RSMM and RSMM-MR demonstrate that the multiresolutional architecture indeed helps to improve the performance. The averages of AUC and pAUC increase by 0.0348 and 0.0757 after incorporating multiresolutional architecture in RGMM and RSMM, respectively. % .  

\subsection{Effect of a temporal attention mechanism}
Lastly, we investigate the effect of the temporal attention mechanism described in Section \ref{TAM}. % . 
As shown in Table \ref{table3} and Figs. \ref{fig5} and \ref{fig6}, the results of RSMM-MR without the attention mechanism are consistently lower than RSMM-MR. This demonstrates that the recurrent layer augmented with attention mechanism is more effective to capture temporal dynamic in time series data.

\section{Robustness of the model to noisy training data}\label{robustness}
\begin{figure}[h]
	\centering
	\includegraphics[width=0.4\columnwidth]{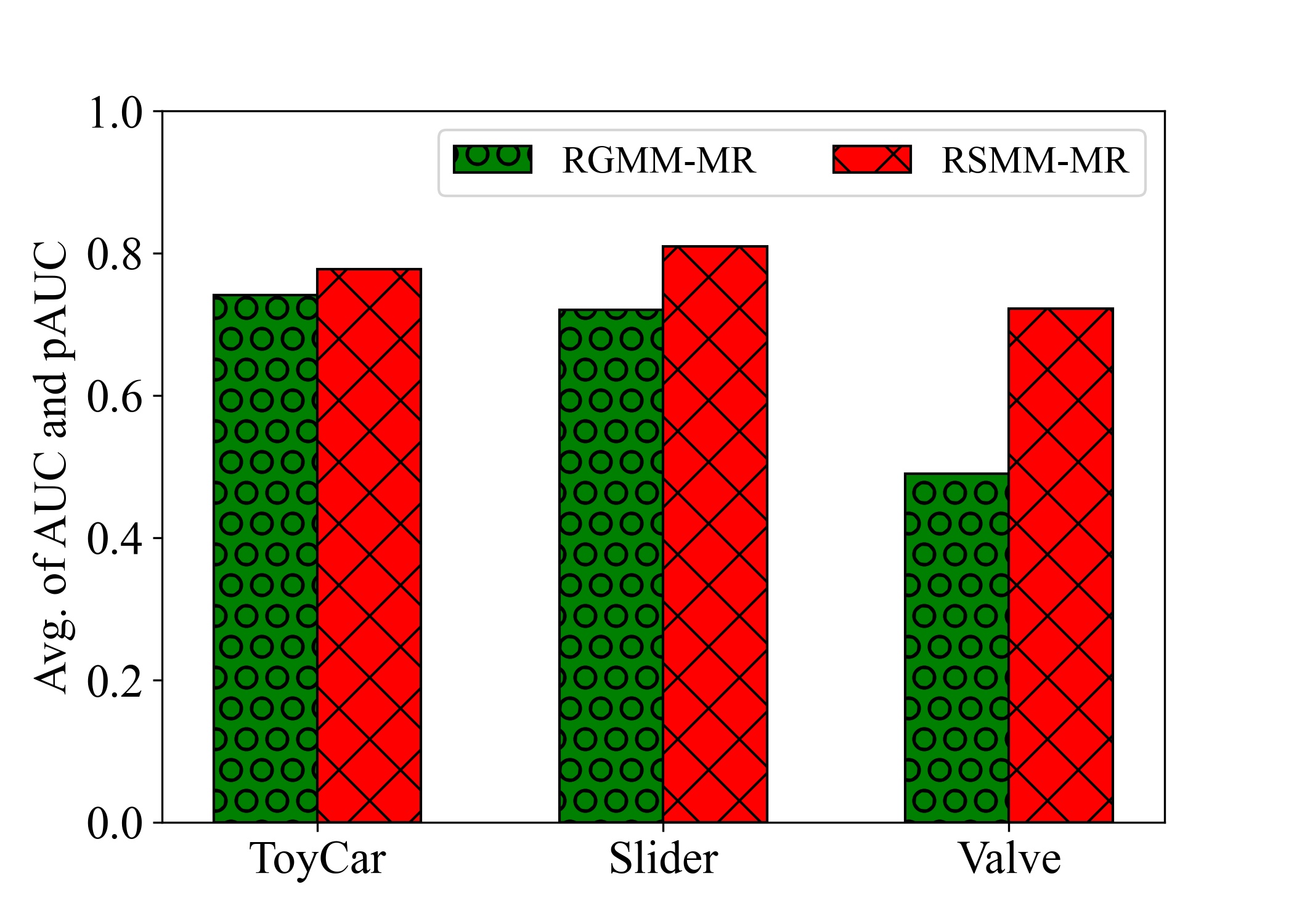} %.
	\caption{Results of introducing Gaussian noise bursts to training data}
	\label{fig7}
\end{figure}
To further investigate the robustness of the proposed model, we conduct one more experiment using ToyCar (\texttt{ToyADMOS}), Slider (\texttt{MIMII}), and Valve (\texttt{MIMII}). The effectiveness of an outliers robust distribution modeling the innovation process is highlighted by randomly introducing bursts of zero-mean Gaussian noise with variance $\sigma^2$ = 5 to 10\% of the overall processing frames in spectrogram. The result is summarized in Fig. \ref{fig7}, and it shows the average values of AUC and pAUC of three IDs. As expected, the synthetically induced outliers negatively affect the detection accuracy for both models. However, the performance of our proposed robust model (RSMM-MR) is consistently higher than the non-robust model (RGMM-MR); the average values of AUC and pAUC are higher for Toycar, Slider, and Valve by 0.0366, 0.0986, and 0.2247.
\end{document}